\documentclass{emulateapj}
\usepackage{natbib}

\shortauthors{R. E. Louis et al.}

\begin{document}

\title{Supersonic Downflows at the Umbra-Penumbra Boundary of Sunspots}

\author{Rohan E. Louis\altaffilmark{1}, Luis R. Bellot Rubio\altaffilmark{2},
Shibu K. Mathew\altaffilmark{1} and P. Venkatakrishnan\altaffilmark{1}}

\altaffiltext{1}{Udaipur Solar Observatory, Physical Research Laboratory
                     Dewali, Badi Road, Udaipur,
	       	     Rajasthan - 313004, India}

\altaffiltext{2}{Instituto de Astrof\'{\i}sica de Andaluc\'{\i}a (CSIC),
                     Apartado de Correos 3004,
                     18080 Granada, Spain}

\email{eugene@prl.res.in}

\begin{abstract}
  High resolution spectropolarimetric observations of 3 sunspots taken
  with {\em Hinode} demonstrate the existence of supersonic downflows
  at or close to the umbra-penumbra boundary which have not been reported
  before. These downflows are confined to large patches, usually encompassing 
  bright penumbral filaments, and have lifetimes of more than 14 hr. The presence of
  strong downflows in the center-side penumbra near the umbra rules
  out an association with the Evershed flow.  Chromospheric
  filtergrams acquired close to the time of the spectropolarimetric
  measurements show large, strong, and long-lived brightenings in the
  neighborhood of the downflows. The photospheric intensity also
  exhibit persistent brightenings comparable to the quiet Sun.
  Interestingly, the orientation of the penumbral filaments at the
  site of the downflows is similar to that resulting from the
  reconnection process described by \citet{Ryutova2008a}.
  The existence of such downflows in the inner penumbra represents a challenge for
  numerical models of sunspots because they have to explain them in terms of physical 
  processes likely affecting the chromosphere.
\end{abstract}

\keywords{Sun: magnetic fields---sunspots---techniques: polarimetric}

\section{Introduction}
\label{intro}
The Evershed flow \citep[EF;][]{Evershed1909} is a well known
phenomenon associated with the filamentary structure of sunspot
penumbrae \citep[][and references therein]{Solanki2003}. The EF is
observed as a shift of the spectral lines, due to a nearly horizontal,
radial outflow of plasma that starts as upflows in the deep layers of
the inner penumbra \citep{Luis2006,Rimmele2006,Franz2009} and ends in
a ring of downflow channels in the mid and outer penumbra
\citep{Westendorp1997,Schlichenmaier1999,Shibu2003,Luis2004,Borrero2005}.
The weak Evershed upflows in the inner penumbra are associated
with bright penumbral grains that migrate towards the umbra-penumbra
boundary with speeds of $\approx$ 1 km~s$^{-1}$ \citep{Rimmele2006}
while individual velocity packets are observed to propagate towards
the periphery of the sunspot with velocities of 2--5.5 km~s$^{-1}$
\citep{Rimmele1994} and vary on a time scale of 10--20 min
\citep{Rimmele1994,Shine1994,Rouppe2003,Cabrera2007}.

Apart from the EF, the penumbra harbors other types of mass motions.
Recent {\em Hinode} observations by \citet{Jurcak2010} have revealed
localized downflows that do not appear to be related to the EF.  These
downflows, with velocities of up to 1 km~s$^{-1}$, have a typical size
of 0\farcs5.  Some of them are seen to be co-spatial with Ca
brightenings in the chromosphere.  \citet{Jurcak2010} pointed out that
the downflowing areas have the same polarity as the parent sunspot and
are unlikely to be the Evershed mass returning to the photosphere,
which sometimes happens well within the penumbra
\citep{Luis2004,Luis2007,Alberto2008}.
 
In this paper we describe a new type of downflows that are observed
close to or at the umbra-penumbra boundary and show supersonic
velocities.  They have the same polarity as the spot, occupy an area
larger than 1.5 arcsec$^2$, and occur along bright penumbral filaments.
These properties are different from those of the radial Evershed flow
and the downflows reported by \citet{Jurcak2010}, which indicates a
different physical origin. The strong downflows we have detected 
possibly represent dynamic processes occurring in the penumbra whose 
manifestation is also seen in the overlying chromosphere.

\begin{figure*}
\hspace{20pt}
\centerline{
\includegraphics[angle=90,width=0.97\textwidth]{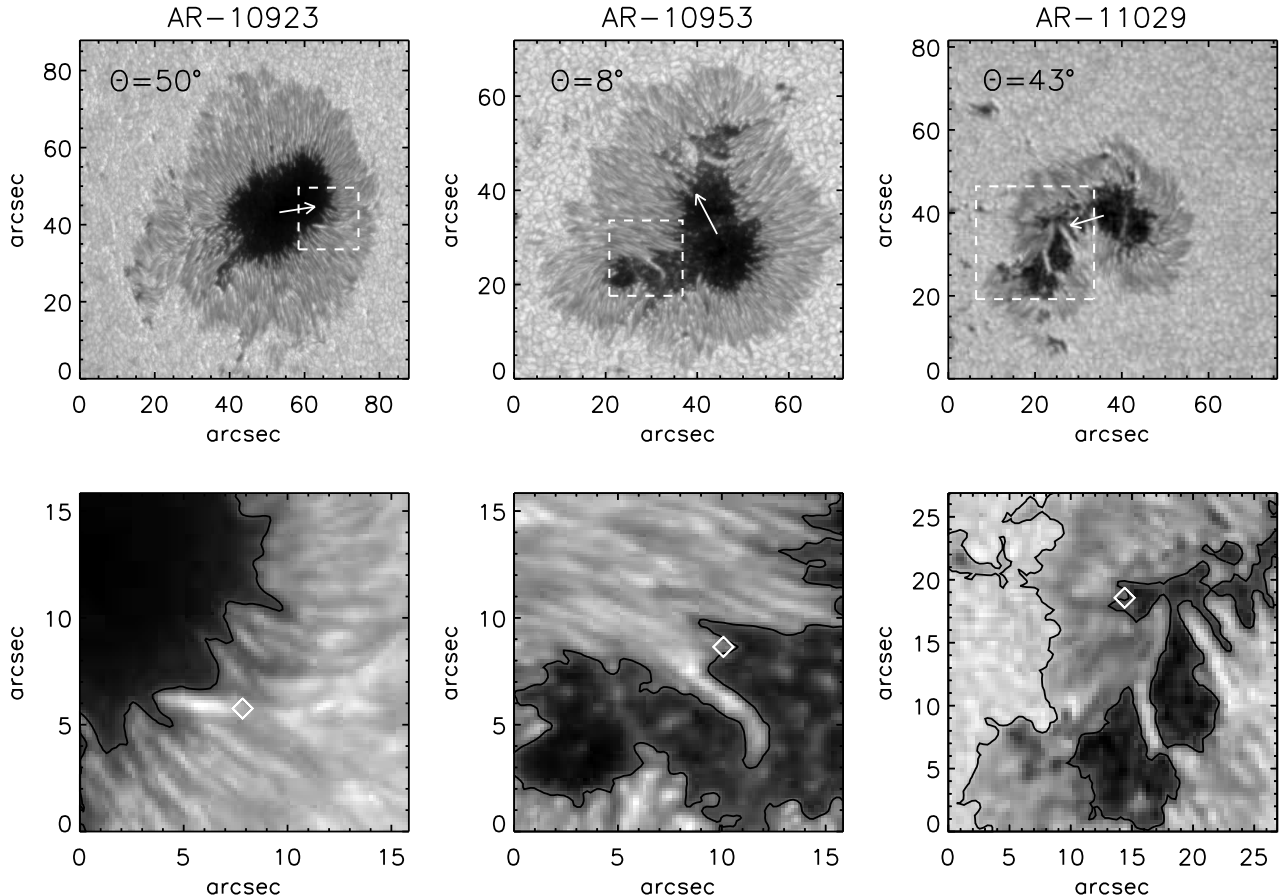}
}
\caption{Sunpots studied in this paper. {\bf Top:} continuum intensity maps
  constructed from the SP scans. The arrow points to disc center. {\bf
    Bottom:} Enlargement of the regions marked by white-dashed
  squares in the top panels. The diamonds represent the pixels whose
  Stokes profiles are shown in Figure~\ref{combo}.}
\vspace{10pt}
\label{cont_image}
\end{figure*}

\section{Observations}
\label{data}

Spectro-polarimetric observations of NOAA AR 10923, 10953 and 11029
were carried out using the Solar Optical Telescope
\citep[SOT;][]{Tsuneta2008} on board {\em Hinode} \citep{Kosugi2007},
on 2006 November 10, 2007 May 1 and 2009 October 27 respectively. The
sunspots were located at heliocentric angles of 50$^\circ$, 8$^\circ$
and 43$^\circ$ as shown in Figure~\ref{cont_image}. They were mapped
by the spectro-polarimeter \citep[SP;][]{Lites2001,Ichimoto2008} from
16:01 UT to 17:25 UT, 21:00 UT to 22:24 UT and 14:45 UT to 15:15 UT
respectively. At each slit position the four Stokes profiles of the
neutral iron lines at 630 nm were recorded with a spectral sampling of
21.55 m\AA. While the first two ARs were scanned in the normal mapping
mode with an exposure time of 4.8~s and a pixel size of 0\farcs16, the
fast mode was used to observe AR 11029 with exposure times of 1.6~s
and a pixel size of 0\farcs32.  The measurements were corrected for
dark current, flat field, thermal flexures, and instrumental
polarization using the SolarSoft package. The zero point of the velocity scale
was set at the line core position of the average quiet Sun profile,
determined considering all the pixels with polarization signals
smaller than three times the noise level. 

G-band and Ca {\sc ii} H filtergrams acquired by he Broadband Filter
Imager (BFI) close to the SP scans were also employed.  The
filtergrams had a sampling of 0.$''$055, 0.$''$11 and 0.$''$11 and a
cadence of 30 s, 1 min and 5 min respectively. These data were
recorded from 13:00 to 14:00 UT, 20:00 to 20:58 UT and 14:00 to 15:55
UT respectively. The Ca images were obtained simultaneously with the
SP scans for AR 11029 while the time separation between the Ca
data set and the Stokes spectra for AR 10923 and 10953 were 3 hr
and 1 hr respectively. The filtergrams were flat fielded and corrected
for dark current and bad pixels. The analysis presented in this paper
pertains to a small sub-region of the center-side penumbra as shown in
the bottom panels of Figure~\ref{cont_image}.

\begin{figure*}
\centerline{
\hspace{30pt}
\includegraphics[angle=90,width=0.95\textwidth]{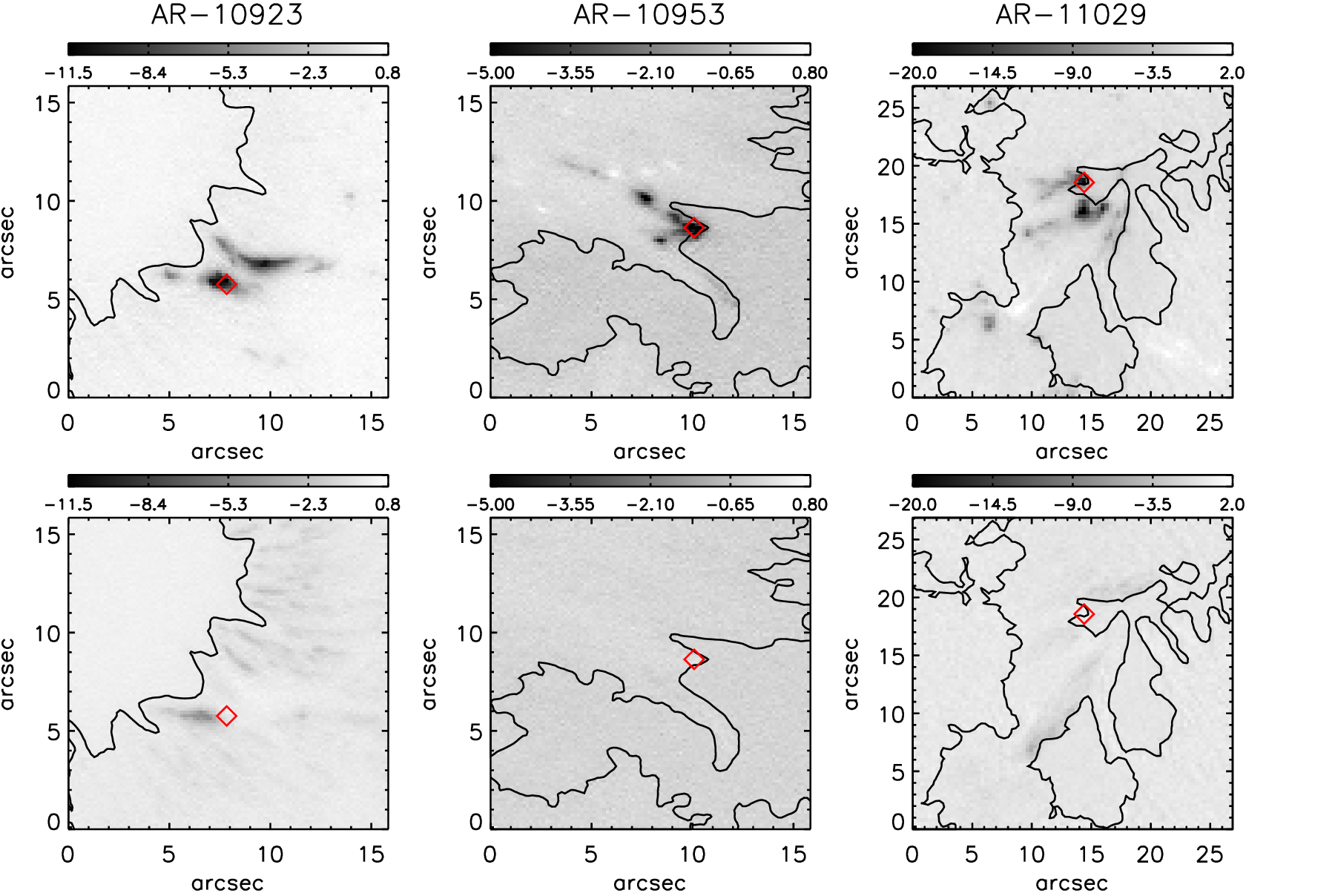}
}
\vspace{-30pt}
\centerline{
\hspace{30pt}
\includegraphics[angle=90,width=0.95\textwidth]{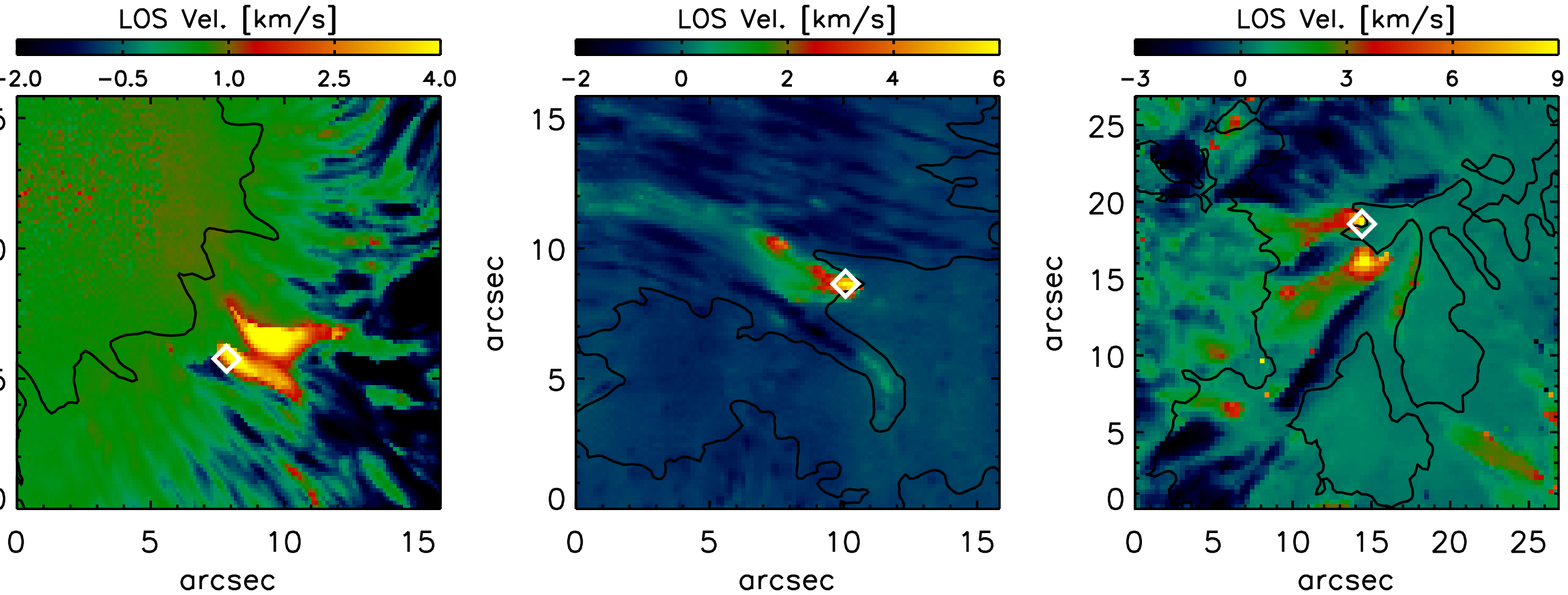}
}
\vspace{-135pt}
\caption{{\bf{Top: }}Red wing magnetograms constructed at $+$34.4 pm
  from the 6302.5\AA~line for NOAA ARs 10923, 10953 and 11029 (from
  left to right respectively). The sign of the signal has been
  reversed. {\bf{Center: }}Blue wing magnetograms constructed at
  $-$34.4 pm from the 6302.5\AA~line. The numbers in the gray scale
  color bars are expressed in per cent. {\bf{Bottom: }}LOS velocities
  derived from the SIR inversion. Positive velocities indicate
  downflows. The maps have been scaled individually. The diamonds
  represent pixels located in strong downflowing regions; their Stokes
  profiles are displayed in Figure~\ref{combo}.}
\label{velo}
\end{figure*}

\begin{figure}
\centerline{
\includegraphics[angle=0,width=0.48\textwidth]{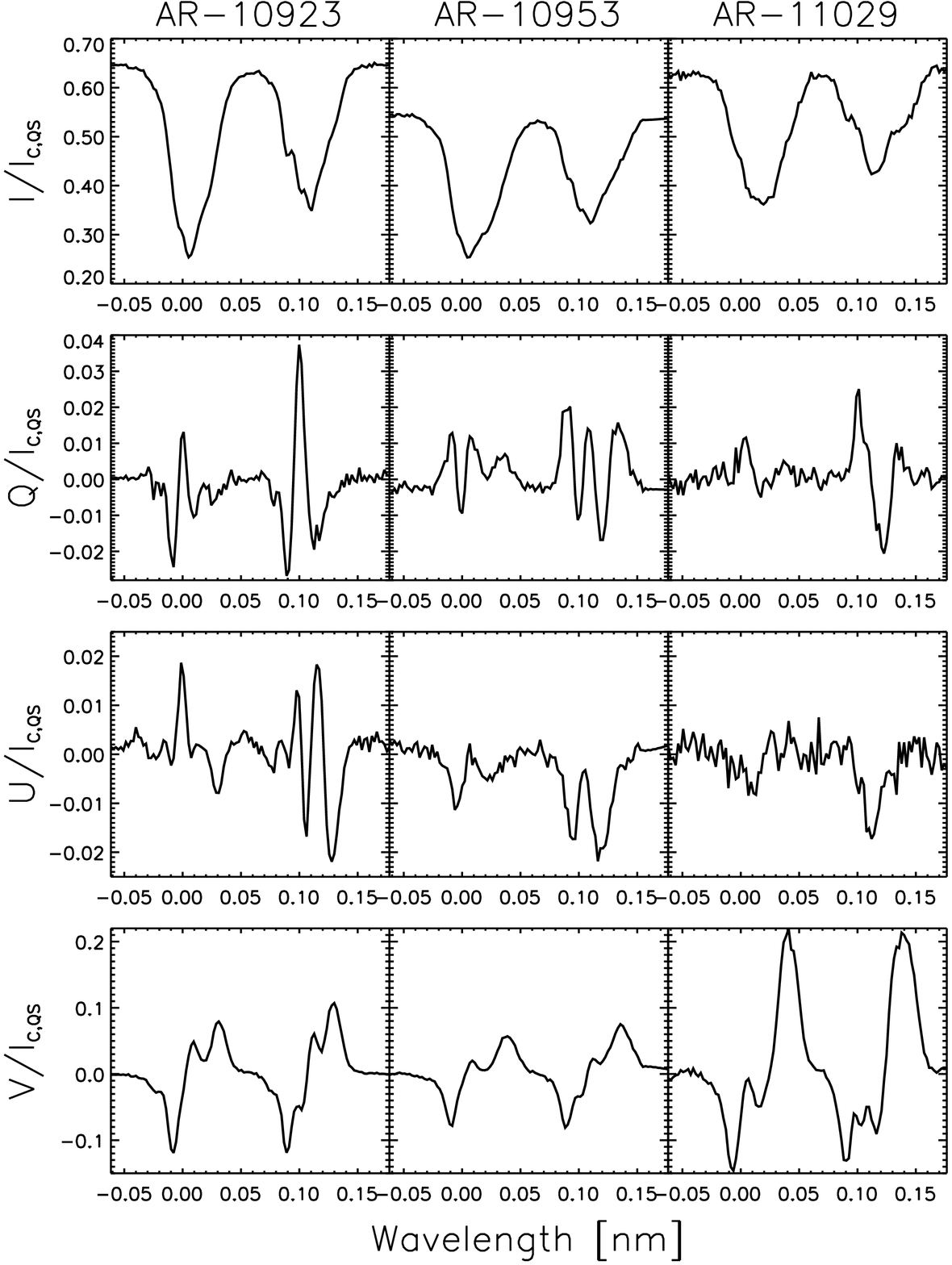}
}
\vspace{-5pt}
\caption{Stokes profiles corresponding to the pixels marked in
  Figures~\ref{cont_image} and \ref{velo}.}
\label{combo}
\end{figure}

\section{Results}
\label{res}
\subsection{Supersonic Downflows}
\label{super}
\subsubsection{Far Wing Magnetograms}
\label{magneto}
To identify sites with large velocities we construct magnetograms in
the far blue and red wings of the Stokes V profile, $\pm$ 34.4 pm from
the center of the 630.25 nm line. This method was applied by 
\citet{Ichimoto2007} to locate the sources and sinks of the Evershed
flow. The sunspots of all 3 ARs have negative polarity and the sign
of the red wing magnetograms has been reversed to match that of the
blue wing.

The top and middle panels of Figure~\ref{velo} display the red and
blue wing magnetograms, both of which have been scaled identically. As
is evident, the red wing magnetograms show very strong signals while
their blue counterparts do not.  In AR 10923 we can see two distinct
filamentary structures very close to one another, with Stokes V
signals of about 9\% (Figure~\ref{velo}, top left panel).  The blue
magnetogram of the same sub-region shows a similar (but weaker)
isolated filamentary structure that lies very close to the
umbra-penumbra boundary. Part of this feature is observed also in the
red wing magnetogram (at the position of the diamond marked in
Figure~\ref{velo}), which is indicative of strong magnetic fields. The
extended feature is also co-spatial with an extremely bright filament
in the corresponding continuum image (Figure~\ref{cont_image}).

Two patches of very strong V signal, separated by $\sim$3\arcsec, are
observed in the red wing magnetogram of AR 10953 with one of them
lying at the umbra-penumbra boundary. In AR 11029 there are two
extended regions near the edge of the umbra, with strong V signals of
up to 20\% of the continuum intensity.

\subsubsection{LOS Velocities from SIR}
\label{sir}
While the far wing magnetograms are useful to locate pixels with large
velocities, the actual magnitude cannot be quantified merely from the
strength of the signals. In order to estimate the range of velocities
present in those pixels, the observed Stokes profiles were subject to
an inversion using the SIR code \citep[Stokes Inversion based on Response Functions;][]{Ruiz1992}. The simplest model
atmosphere was assumed wherein the vector magnetic field (field
strength, inclination and azimuth) and LOS velocity were kept constant
with height, while the temperature was perturbed with 2 nodes. The
inversions also retrieved height-independent micro- and
macro-turbulent velocities as well as the fraction of stray light in
each pixel. 

The lower panels of Figure~\ref{velo} depict the resulting LOS
velocities for the 3 ARs. These values must be regarded as approximate
since large gradients of the atmospheric parameters may exist along
the LOS. The structure of the downflows is remarkably, but not
surprisingly, similar to that seen in the far red wing magnetograms.
While strong downflows of $\sim$5 km~s$^{-1}$ are observed in AR
10923, one finds velocities of around 6 km~s$^{-1}$ and 10 km~s$^{-1}$
in ARs 10953 and 11029 respectively.  The strong downflowing zones are
surrounded by upflows which can be identified with the Evershed flow.

The majority of the Stokes V profile emerging from the downflows
exhibit a satellite or even an additional red lobe. The Stokes $I$
profiles also have highly inclined red wings, as illustrated in the
top row of Figure~\ref{combo} for the pixels marked with diamonds in
Figure~\ref{velo}.  In AR 11029, some of the downflows exhibit
strongly redshifted V profiles with large area asymmetries (bottom row
of Figure~\ref{combo}). Upflows of about 2 km~s$^{-1}$ corresponding
to the EF are seen adjacent to the strong downflowing patches in all 3
ARs.

\begin{figure}
\centerline{
\hspace{120pt}
\includegraphics[angle=90,width=0.425\textwidth]{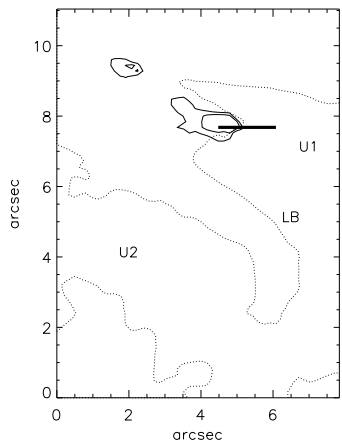}
}
\vspace{10pt}
%\caption{Continuum intensity contour of AR 10953 (black). Red contours
%  mark LOS velocities of 2.5 and 4 km~s$^{-1}$. The blue line
%  represents a cut across the strong downflows at the umbra-penumbra
%  boundary. The distance along the slit increases from left to right
%  with the umbra-penumbra boundary being zero. Pixels to the left are
%  negative. U1, U2 - umbrae; LB - light bridge.}
\caption{Continuum intensity contour of AR 10953 (black dotted lines). Black solid contours
  mark LOS velocities of 2.5 and 4 km~s$^{-1}$. The thick line
  represents a cut across the strong downflows at the umbra-penumbra
  boundary. The distance along the slit increases from left to right
  with the umbra-penumbra boundary being zero. Pixels to the left are
  negative. U1, U2 - umbrae; LB - light bridge.}
\label{con_super}
\end{figure}

\begin{figure*}
\centerline{
\includegraphics[width=0.55\textwidth,angle=90]{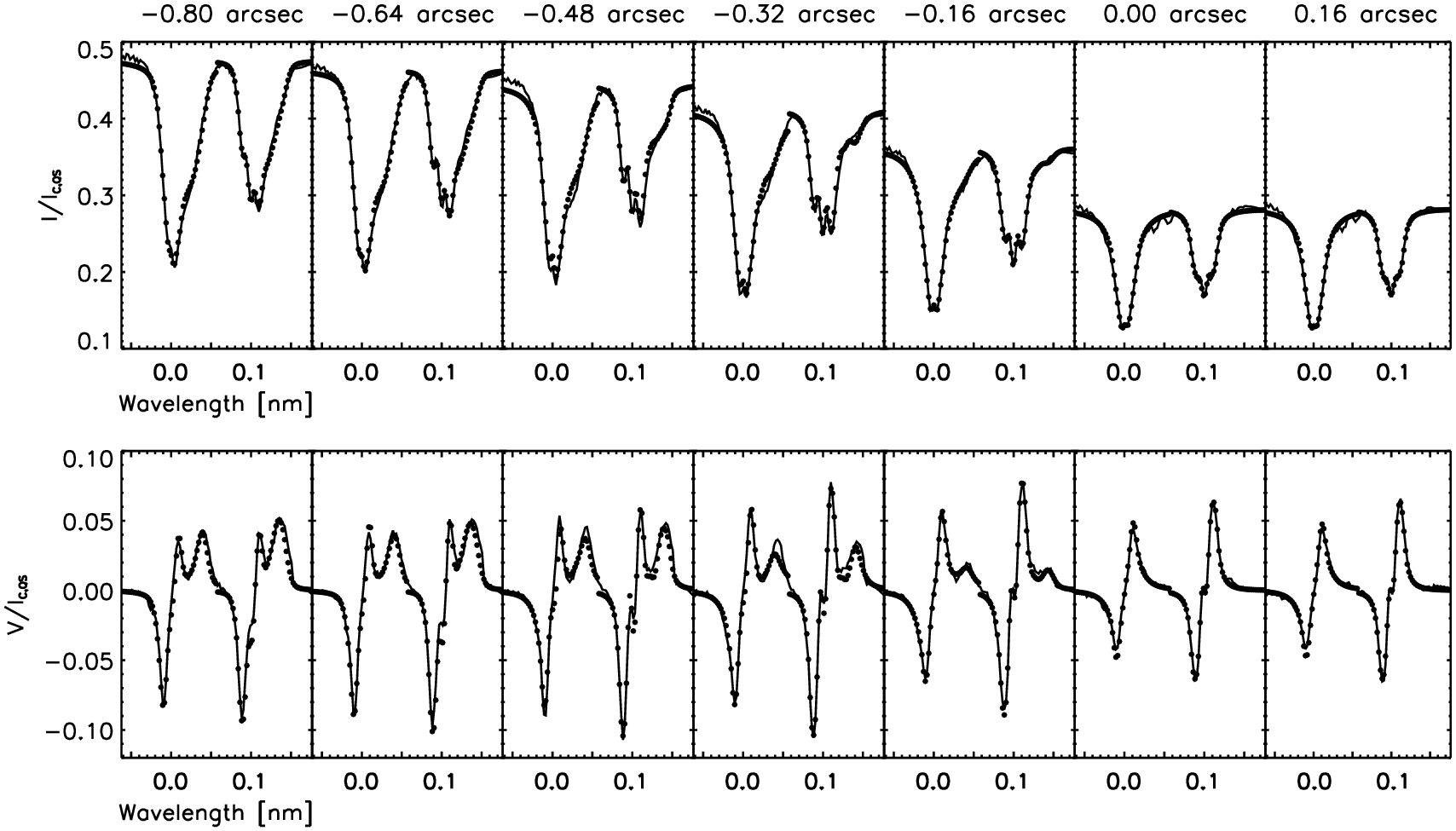}
\vspace{-50pt}
}
\centerline{
\includegraphics[width=0.55\textwidth,angle=90]{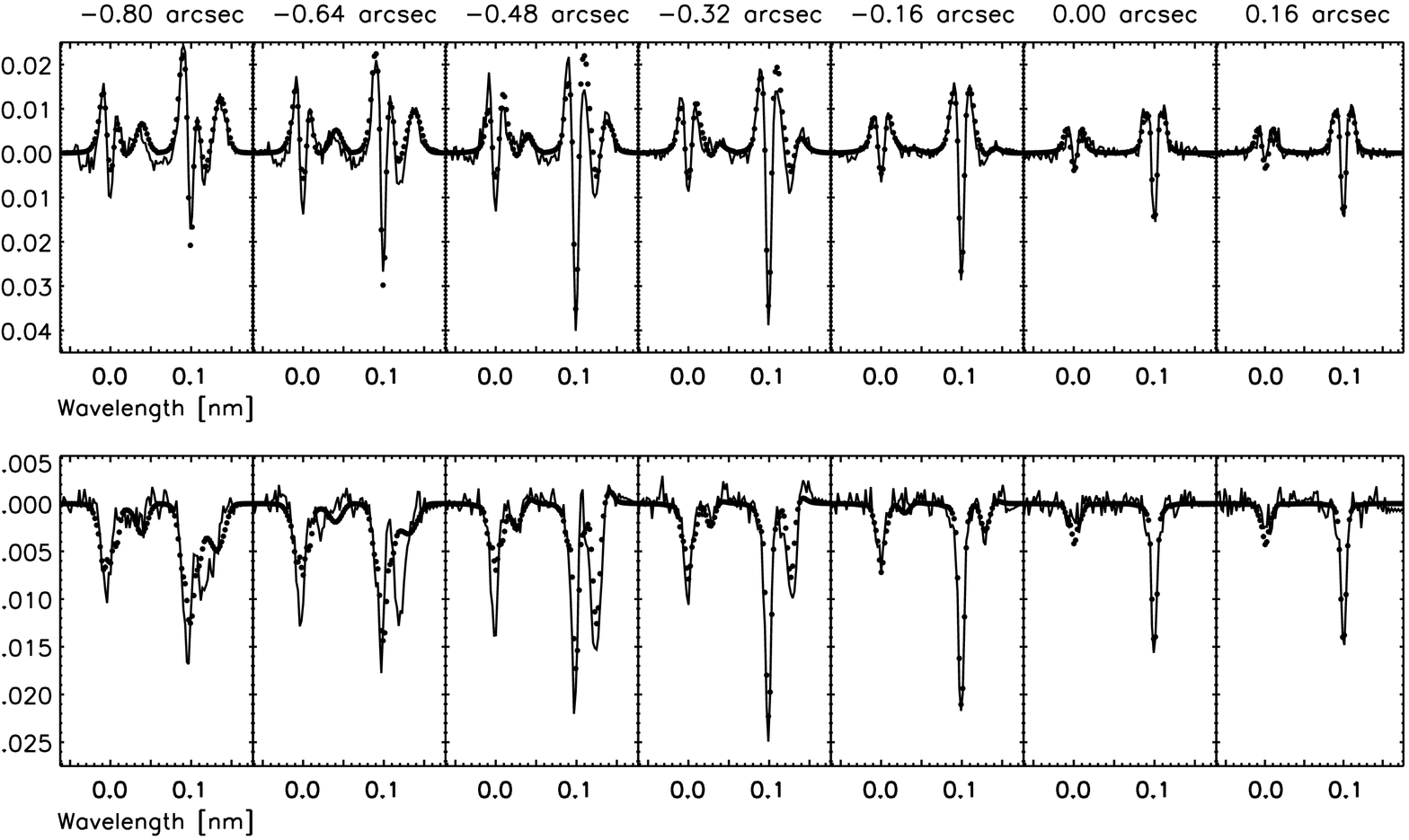}
} 
\vspace{-40pt}
%\caption{Observed ({\em black}) and best-fit ({\em red}) Stokes
%  profiles along the cut marked in Figure 5. From top to bottom:
%  Stokes I, V, Q and U.}
\caption{Observed ({\em black}) and best-fit ({\em black filled circles}) Stokes
  profiles along the cut marked in Figure 5. From top to bottom:
  Stokes I, V, Q and U.}

\label{umbra_stokes}
\end{figure*}

\begin{figure}
\centerline{
\includegraphics[angle=90,width=0.45\textwidth]{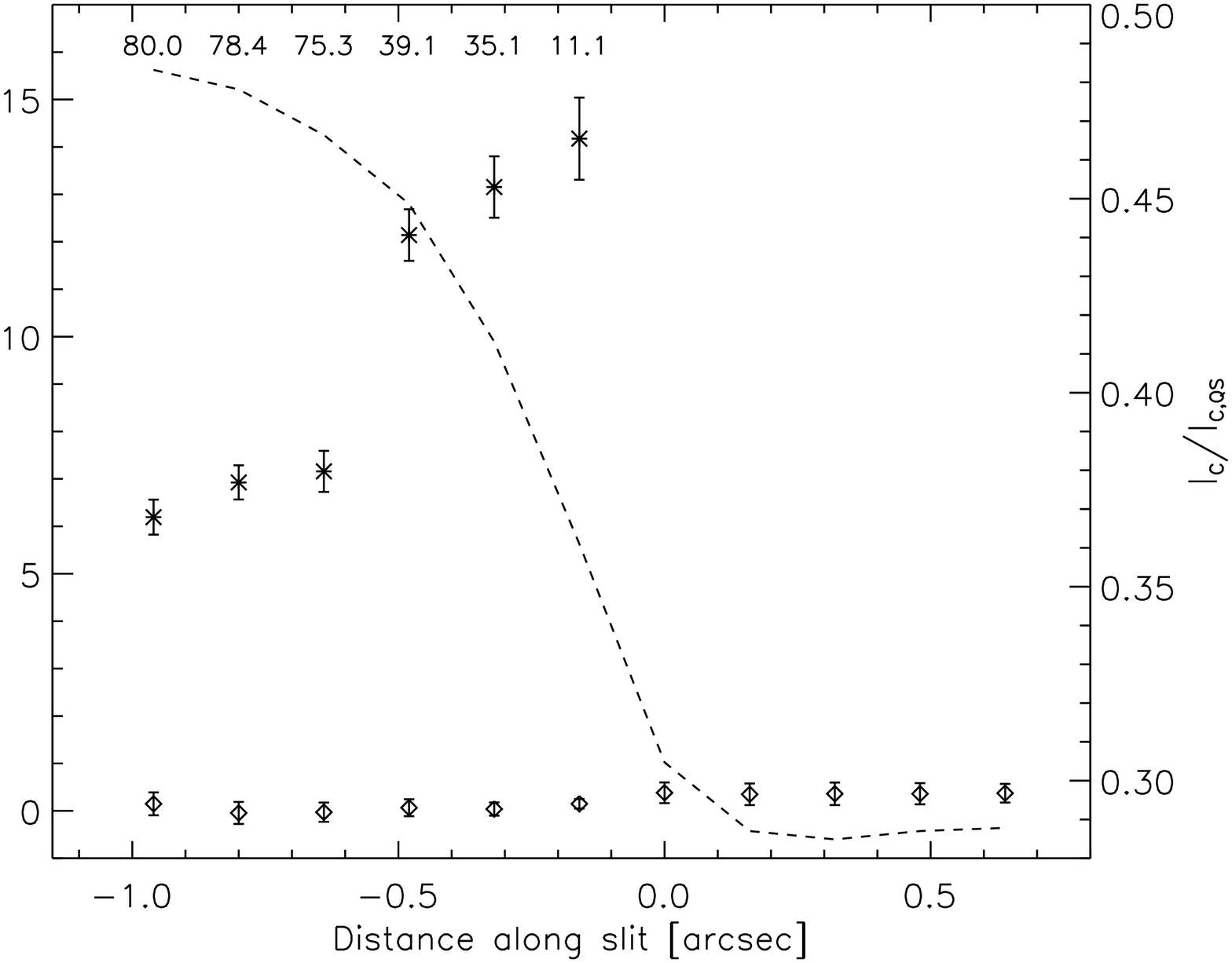}
}
\vspace{20pt}
\caption{Variation of LOS velocity along the cut displayed in Figure
  4, as resulting from an inversion of the observed Stokes profiles.
  The pixels left of the umbra-penumbra boundary were inverted using a
  two-component model atmosphere. The diamonds and asterisks represent
  the slow and fast component, respectively. The error bars correspond
  to $\pm$ 1 sigma. The dashed line shows the normalized continuum
  intensity along the slit. The numbers below the upper x-axis
  represent the filling fraction of the faster moving component.}
\label{velo_super}
\end{figure}

Figure~\ref{con_super} shows a cut passing through the strong
downflowing patch located at the umbra-penumbra boundary of AR 10953.
The variation of the Stokes profiles along this cut is displayed in
Figure~\ref{umbra_stokes} (black lines). At the umbral edge and to the
right, one observes a distinct reduction in the continuum intensity.
To the left of the umbra-penumbra boundary, the Stokes V profile has
two distinct red lobes while the $I$ profile exhibits highly inclined
red wings. With five lobes, the Stokes Q profile is very asymmetric,
0\farcs32 beyond the edge of the umbra.  The anomalous profiles lying
to the left of the umbra-penumbra boundary were inverted using a 2
component atmosphere and setting all parameters, except the temperature,
to be constant with height in order to estimate the physical
parameters more precisely. The highly inclined red wings of the Stokes
$I$ profiles as well as the double-red-lobed V profiles are well
reproduced by the simple two-component atmosphere (lines with black filled circles in
Figure~\ref{umbra_stokes}). The transition from supersonic to nearly
zero velocities along the cut can be seen in Figure~\ref{velo_super},
which shows that very large velocities nearly twice the sound speed
occur adjacent/close to the umbra-penumbra boundary and measure
$\approx$ 7.2 km~s$^{-1}$ at a distance of $-0\farcs64$. At the umbral edge 
the faster component has a fill fraction of $\approx$ 11\% which increases 
with distance and beyond $-0\farcs64$ fills more than 70\% of the pixel. 
While the fast and slow components appear to differ by $\approx$ 300 G
the uncertainties in the retrieved parameters suggest that the magnitudes 
are nearly the same. Both components have field strengths in excess of 2.2
kG. The average zenith angle in the umbra is $\approx$ 143$^\circ$ and
changes from 142$\pm$2$^\circ$ at the umbra penumbra boundary to
137$\pm$5$^\circ$ at $-0\farcs96$ for the slow component. In comparison, 
the zenith angle of the fast component varies from 150$\pm$11$^\circ$ to 
124$\pm$4$^\circ$ for the same spatial locations.

The typical area of one downflowing patch varies from 1.6 arcsec$^2$ 
for AR 10953 to as large as 6 arcsec$^2$ as observed in
AR 11029. The two components could reside side-by-side in the same
resolution element or could be stacked one on top of the other in the
vertical direction. Regardless of the exact configuration, supersonic
velocities exist in the presence of very strong magnetic fields. In
addition, the polarity of the strong downflowing component is the same
as that of the sunspot, which rules out the possibility of them being
Evershed flows returning to the solar surface. The strong fields would
also inhibit convection, hence the downflows appear to be unrelated to
the Evershed phenomenon and are likely to be caused by an alternative
mechanism.

\subsection{Chromospheric Activity}
\label{ca-img}
\begin{figure}
\centerline{
\includegraphics[angle=0,width=0.72\textwidth]{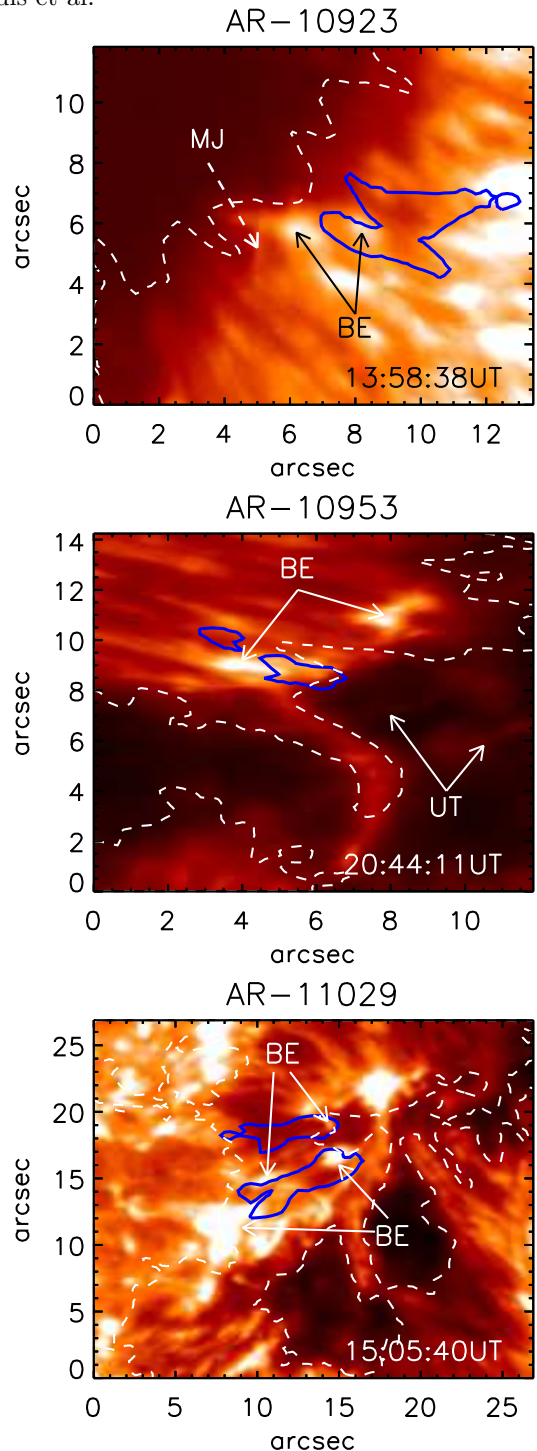}
}
%\caption{Ca {\sc ii} H filtergrams acquired close to the time of the
%  SP scans. Blue contours of LOS velocity greater than 2 km~s$^{-1}$
%  have been overlaid on the filtergrams. The black contour corresponds
%  to the continuum intensity at 630 nm. BE - brightness enhancement;
%  MJ - penumbral mircrojet; UT - bright umbral thread.}
\caption{Ca {\sc ii} H filtergrams acquired close to the time of the
  SP scans. Blue contours of LOS velocity greater than 2 km~s$^{-1}$
  have been overlaid on the filtergrams. The white dashed contour corresponds
  to the continuum intensity at 630 nm. BE - brightness enhancement;
  MJ - penumbral mircrojet; UT - bright umbral thread.}
\label{ca-vel}
\end{figure}

Figure~\ref{ca-vel} depicts Ca {\sc ii} H filtergrams taken close to
or during the time of the SP scans for the 3 ARs. In AR 10923, one can
identify penumbral microjets (MJ) as well as brightness enhancements
(BE) in the vicinity of the strong downflowing areas. The jet is the
one shown in Figure 4 of \citet{Ryutova2008b}. It is oriented nearly
perpendicular to the filament, while the brighteness enhancements
appear as isolated zones on the filaments.  These BEs are $\sim 77$\%
brighter than the jets, which appear to be in the decay phase
\citep{Ryutova2008b}. We will hereafter refer to the brightness enhancements
and microjets as BEs and MJs respectively.

AR 10953 shows a very strong enhancement (BE) in between the two 
downflowing patches that is twice as intense as the neighbouring penumbra. 
Narrow, thread like structures (UT) of reduced intensity in the umbra
are observed to connect to the nearby light bridge.  

Ca filtergrams with a cadence of 5 min taken simultaneously with the
SP scan for AR 11029 exhibit an intense brightening at the
umbra-penumbra boundary within one of the strong velocity patches.
Even stronger brightenings are seen close to the outer penumbral
boundary that lie at the rear end of the extended downflowing patch.
A relatively weaker brightening is seen in the other velocity patch
and occupies a much smaller area. These enhancements are $\sim$ 50 and
15\% brighter than the adjacent penumbral filaments.

\begin{figure}
\centerline{
\includegraphics[angle=0,width=0.7\textwidth]{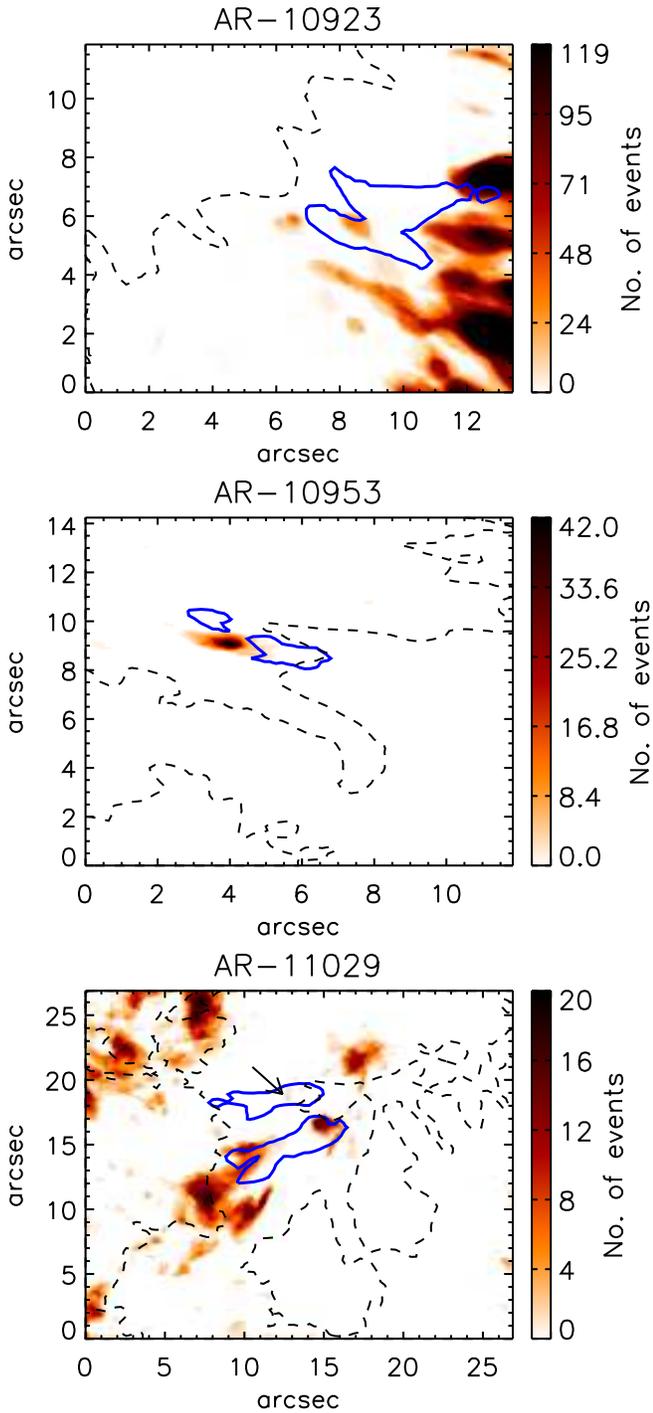}
}
\caption{Ca event maps depicting locations with strong, persistent
  enhancements. The images have been scaled as shown by the vertical
  color bar. The black arrows indicate relatively short lived events.
  See text for details.}
\label{event}
\end{figure}

\begin{figure}
\centerline{
\includegraphics[angle=0,width=0.7\textwidth]{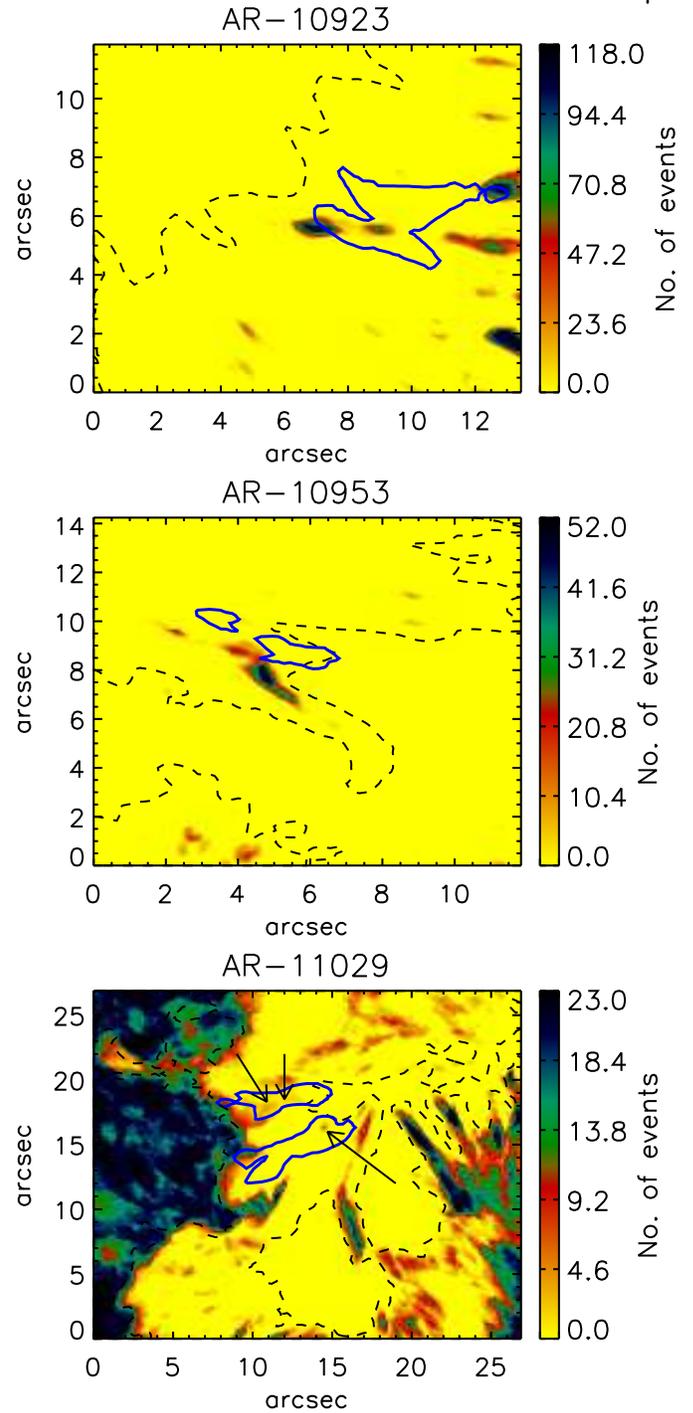}
}
\caption{G-band event maps depicting locations with intensity close to the quiet Sun photosphere. The black arrows in the bottom panel indicate the few strong photospheric brightenings that lie
within the downflowing patches.}
\label{gb_event}
\end{figure}

\begin{figure*}[!t]
\centerline{
\hspace{20pt}
\includegraphics[angle=0,width=0.578\textwidth]{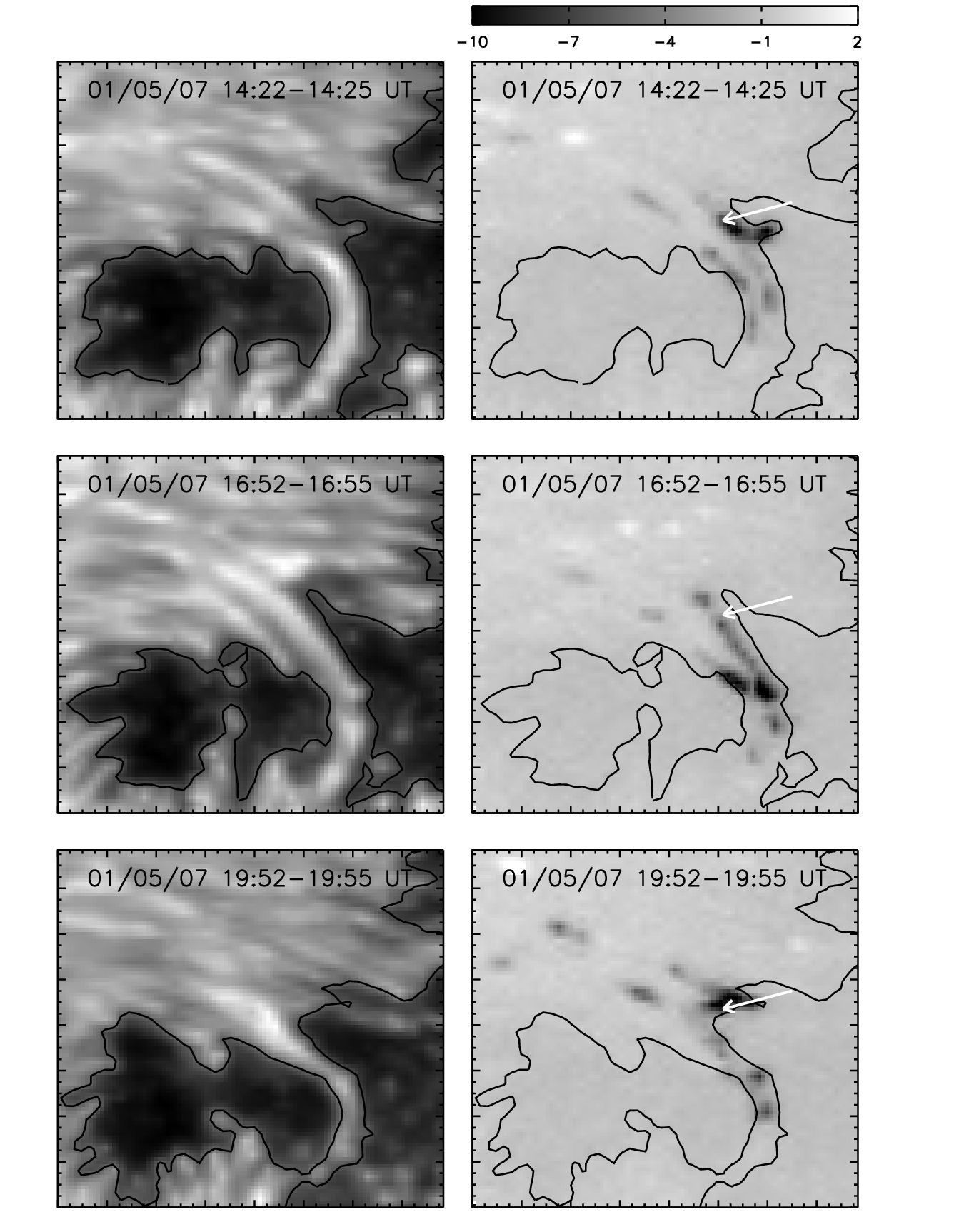}
\hspace{-40pt}
\includegraphics[angle=0,width=0.578\textwidth]{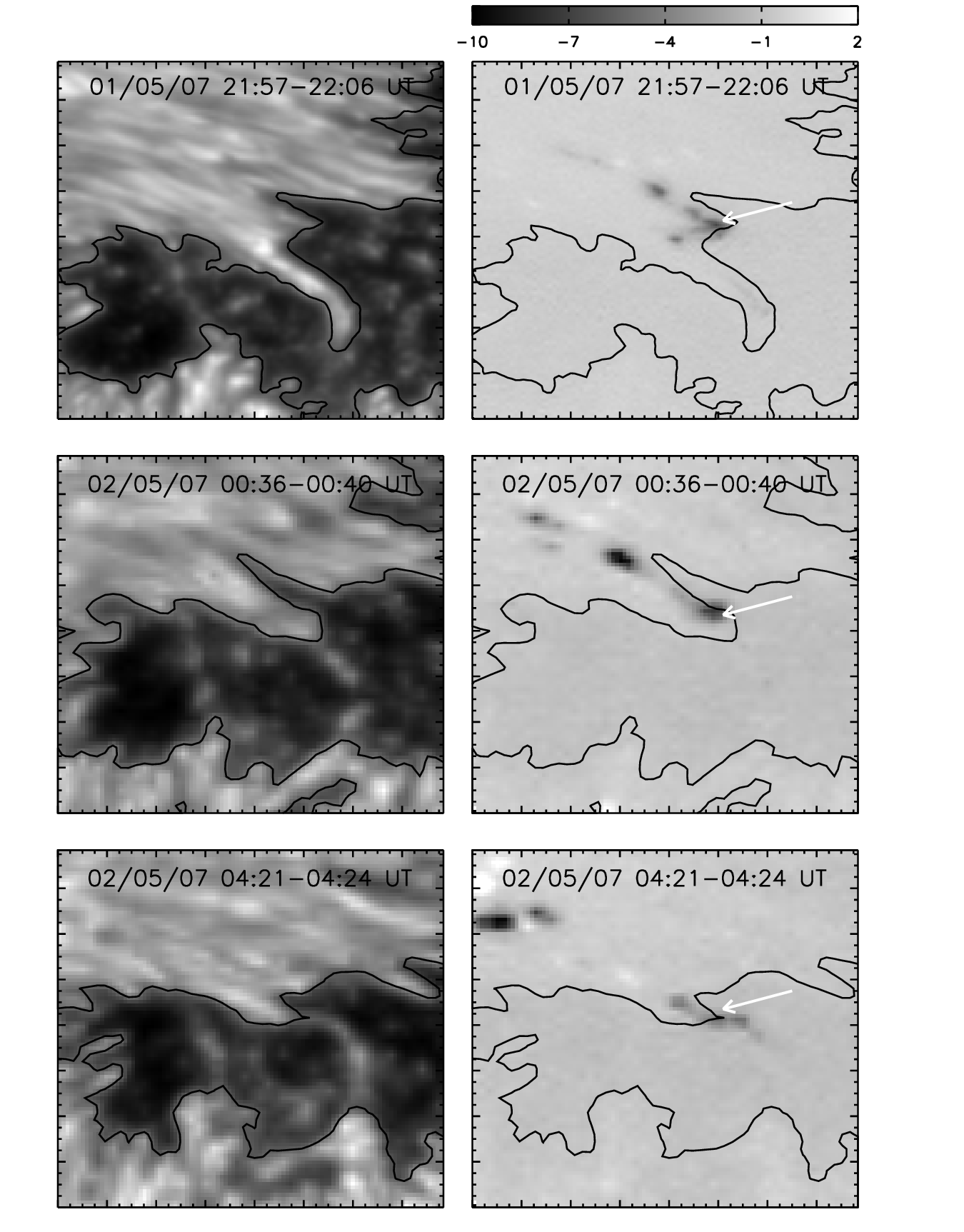}
}
\caption{Temporal evolution of the strong downflows observed in AR
  10953, with continuum images to the left and red wing magnetograms
  to the right.  The date and time of the SP scan is indicated
  in each panel. All the magnetograms have been scaled identically and
  are given in per cent. The minimum value in the color bar has been
  clipped for better visibility of the downflowing regions. Each
  tickmark in the image corresponds to 0\farcs5. The white arrow
  indicates a location corresponding to the downflows that was tracked
  in the magnetogram time sequence.}
\label{down_may}
\end{figure*}

\begin{figure*}[!t]
\centerline{
\hspace{20pt}
\includegraphics[angle=90,width=0.69\textwidth]{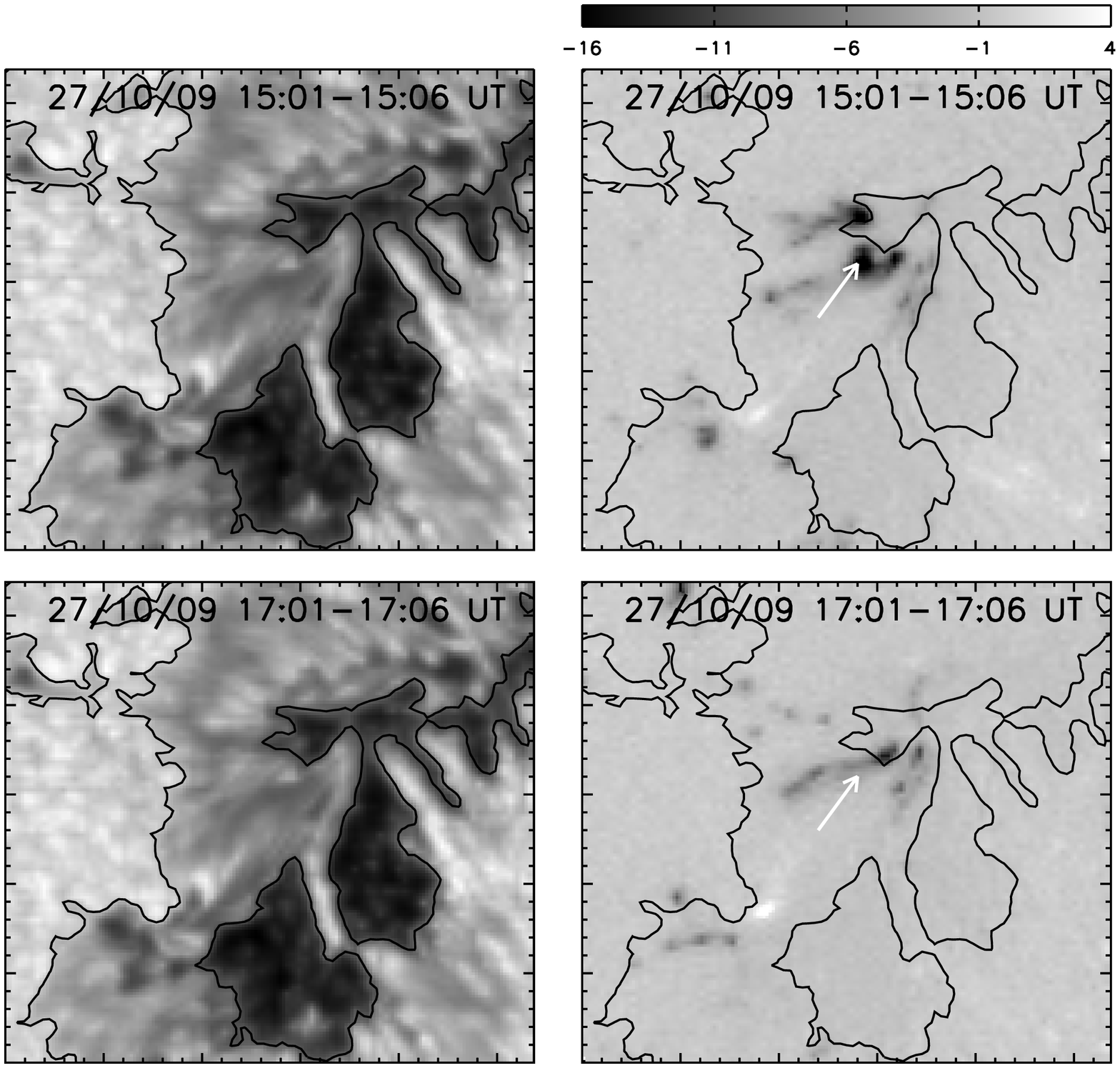}
\hspace{-10pt}
\includegraphics[angle=90,width=0.69\textwidth]{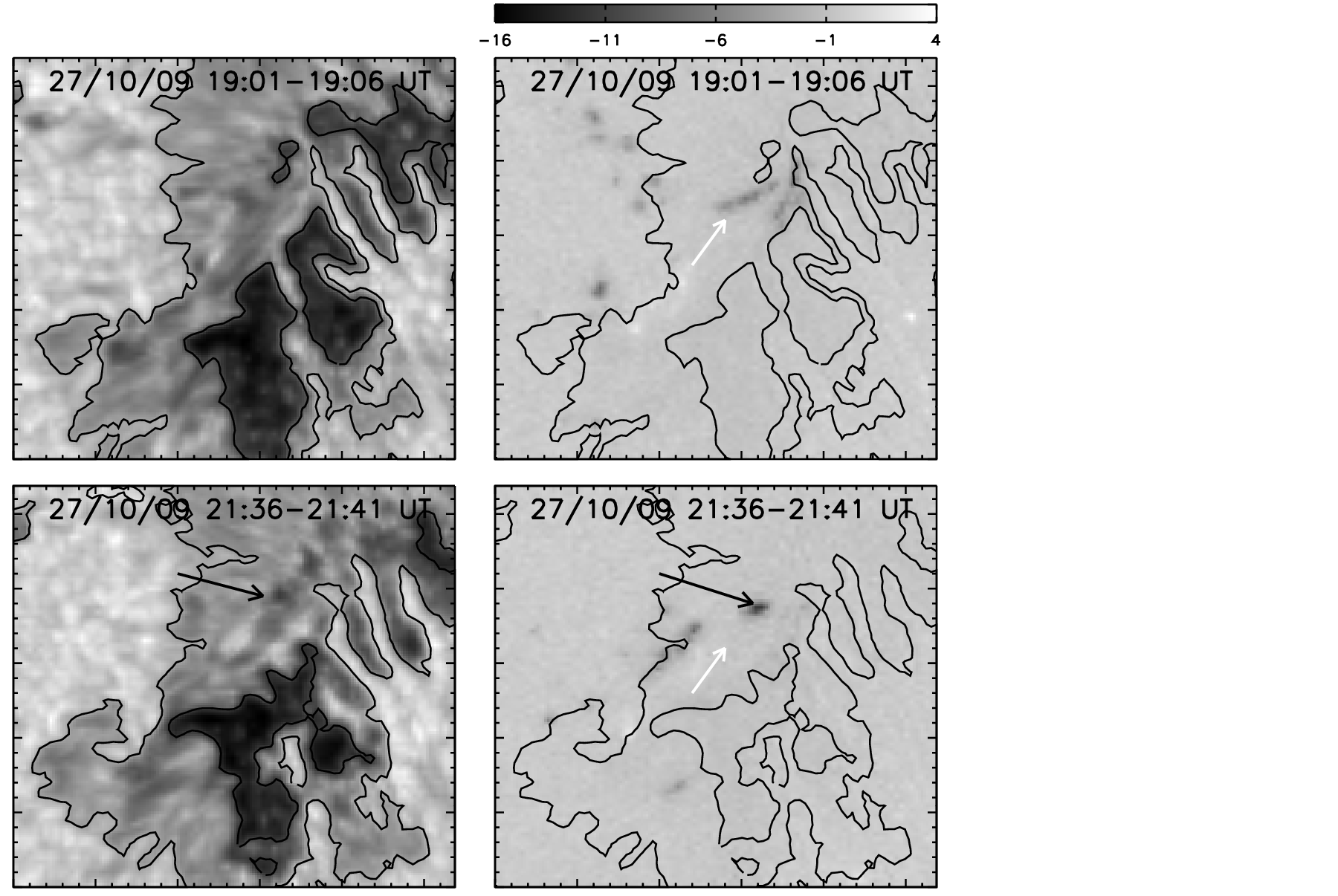}
}
\vspace{10pt}
\caption{Same as Figure~\ref{down_may}, for AR 11029. Each tickmark in
  the images corresponds to 1\arcsec. The black arrow in the bottom
  panels indicate the location of the isolated downflowing patch close
  to the diffuse umbral fragment.}
\label{down_oct}
\end{figure*}

While Figure~\ref{ca-vel} shows single chromospheric snapshots
highlighting the proximity of the downflows and the chromospheric
enhancements, it does not provide information about whether the
spatial correspondence between the two endures with time. In order to
separate the short lived enhancements from the persistent ones, event
maps were constructed from the time sequence of Ca filtergrams
acquired before/during the SP scans in the following manner. First, a
small quiet Sun region was selected to determine the time averaged
chromospheric intensity for each of the sequences. This value was then
used to normalize the intensity of the individual filtergrams. A
histogram of the intensities in the AR was derived to determine a
suitable threshold value. From the trailing part of the histogram,
threshold values of 0.9, 1.0 and 2.0 were chosen for the three ARs
respectively. Using these values, a binary image was created for each
individual image in the time sequence, where all pixels with an
intensity above the threshold were set to one and the rest to zero.
The binary maps were then added in time, yielding at the end, a map
with pixels having values indicating the number of chromospheric
events. The resulting event maps are shown in Figure~\ref{event}.

The map constructed for AR 10923 displays two large patches on the
penumbral filaments at the site of the downflows. They are labeled as
BE in the top panel of Figure~\ref{ca-vel}. The counts at those
locations indicate that brightness enhancements persisted for nearly 
one-third of the 1 hr sequence. There are no obvious signatures of the
MJs, reflecting the transient nature and weakness of these events.

The strongest and long-lived chromospheric brightenings observed in AR
10953 are confined to a region between the two downflowing patches. AR
10953 produced several chromospheric enhancements from April 29 to the
end of May 1 \citep{Rohan2008,Shimizu2009}. Some of them were
co-spatial with supersonic downflows observed in the light 
bridge nearly 10 hours earlier than the downflows reported here
\citep{Rohan2009}. 

The bottom panel of Figure~\ref{event} shows the event map
corresponding to AR 11029. The Ca filtergrams were acquired during the
SP scan with a low cadence of 5 min. Here the strong and persistent
enhancements at the umbra-penumbra boundary are co-spatial with the
largest photospheric downflows. Similar brightenings are seen along
the edges of the downflowing regions near the outer penumbral boundary. 
A relatively shorter but stronger event is observed in the other 
downflowing zone (black arrow in bottom panel of Figure~\ref{event}). 
Several small C class flares were detected in NOAA AR 11029 during 
its passage on the solar disk from October 25 to October 28.

The chromospheric brightenings associated with strong downflowing
patches near the umbra/penumbra boundary are more intense, persist in
time and occupy a large area than the penumbral microjets discovered
by \citet{Katsukawa2007}. Microjets are transient events seen
everywhere in the chromosphere of sunspot penumbrae, with lifetimes of
1 min or less. They show typical widths of 400 km and lengths varying
from 1000 to 4000 km. In contrast, our events have areas of 1--2
arcsec$^2$ and appear as blobs or patches. Their lifetimes are
significantly longer than 1 min.

\subsection{Photospheric Brightness}
\label{photo_down}          
In order to ascertain if the downflows are associated with long lived
photospheric brightenings, G-band event maps were constructed as
described in the previous section.  The same procedure was adopted,
setting threshold values of 0.85, 0.9 and 0.95 of the quiet Sun
intensity for the 3 ARs. Figure~\ref{gb_event} displays the G-band
event maps derived using these threshold values.  The strong, long
lived chromospheric enhancements that were seen in AR 10923 are nearly
co-spatial with G-band brightenings, although the sizes of the
photospheric enhancements are relatively smaller than their
chromospheric counterparts.

The G-band event map of AR 10953 consists of a single isolated blob close 
to the southern edge of the strong downflowing patch located near the 
umbra-penumbra boundary. This blob resembles a comet-like shape with the 
tail end protruding into the nearby light bridge. Relatively fewer,
though stronger events are also observed in between the two downflowing 
patches as well as close to the downflow area located furthest from the 
umbra-penumbra boundary. 

In comparison to ARs 10923 and 10953, the photospheric event map of AR
11029 only shows traces of strong intensity in the downflowing areas
at sporadic instances (see black arrows in the bottom panel of
Figure~\ref{gb_event}.)  If one assumes that the number of
photospheric counts/events observed within the downflowing areas of AR
11029 occurred consecutively, the maximum lifetime of these events
would be equivalent to 35-40 minutes, since the filtergrams were
acquired with a cadence of 5 min. In contrast, the lifetimes of
similar brightenings in the other ARs is close to 1 hr.

The G-band event maps suggest that the process responsible for the
supersonic downflows and the chromospheric enhancements is able to
produce very strong brightenings in the photosphere that are nearly as
intense as the quiet Sun but are also long lasting with lifetimes of
up to 1 hr. Bright penumbral features close to the umbra-penumbra 
boundary with intensities nearly twice that of the quiet Sun have been 
detected by \citet{Denker2008} who speculate that these intermittent 
phenomena may be the outcome of reconnection between adjacent penumbral 
flux tubes.

\subsection{Temporal Evolution of the Downflows}
\label{down_evol}
In this section we study the evolution of the downflows in order to
determine their lifetimes and to establish if they recur or change
their position relative to the umbra-penumbra boundary. To that end, 5
additional data sets were selected for AR 10953 and 3 for AR 11029.
AR 10923 is not considered here because only 2 more scans were
available for analysis. All the observations were calibrated as
explained in Section~\ref{data}, constructing red wing magnetograms 
at 34.4 pm from the 6302.5\AA~line to identify the sites of strong
downflows. In accordance with Figure~\ref{velo}, the sign of the 
red wing magnetogram has been reversed. 

\subsubsection{NOAA AR 10953}
\label{ar_23}
The evolution of AR 10953 over the course of 14 hr from May 1 to May 2
is depicted in Figure~\ref{down_may}. The continuum images and the red
wing magnetograms were co-aligned with the scan taken at 21:57-22:06
UT with an accuracy of $\pm$1 pixel. The white arrow tracks the same
downflowing region in time.  The magnetogram signal at the position of
the arrow is maximum ($\sim$10\%) during the first scan, taken at
14:22-14:25 UT.  In the second scan the signal drops to nearly half
this value before increasing to $\sim$8\% at 19:52-1955 UT.  However,
there are even stronger signals at the umbra-penumbra boundary where
the amplitude of the magnetogram signal is $\sim$12\%.  This is nearly
twice the signal detected in the next frame, which was analyzed
earlier in Section~\ref{magneto}.  The scan made by the SP between
00:36-00:40 UT shows 3 isolated downflowing patches with one of them
located at the umbra-penumbra boundary and a second stronger downflow
area 5\arcsec\/ away from the first. A third patch of downflows with
much weaker signals is seen at a larger distance from the
umbra-penumbra boundary.  The last magnetogram, corresponding to
04:21-04:24 UT, shows diffuse signals of up to 5\% at the
umbra-penumbra boundary.

The time sequence of magnetograms illustrates the following: i) the
strong downflows occurring at the umbra-penumbra boundary continue to
be seen at this location in all 6 SP scans considered here, albeit
with varying strength, which would put a lower limit of about 14 hr on
their lifetimes; ii) the area of the strong downflowing patches at the
umbra-penumbra boundary alone remains fairly constant ($\sim$2
arcsec$^2$) during the latter half of the sequence from 21:57-04:24
UT.

\begin{figure}
\vspace{10pt}
\hspace{15pt}
\centerline{
\includegraphics[angle=0,width=0.8\textwidth]{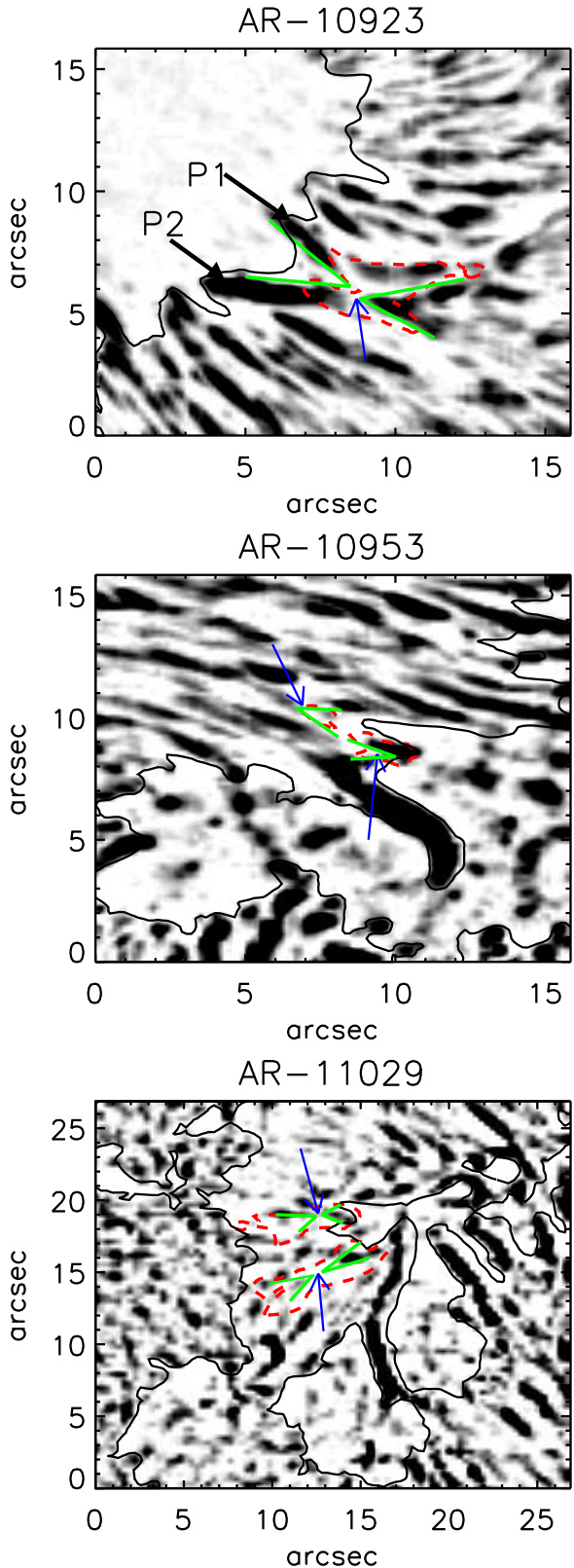}
}
\caption{Morphologically {\em opened} images subtracted from the
  original image to separate bright filaments from the dark
  background, filament crossings, and edges. The images are shown in
  negative.  The dark and bright structures correspond to the bright
  filaments and their diffuse counterparts, respectively. The blue
  arrows indicate locations where the filaments appear to intersect
  each other. P1 and P2 are the two interacting filaments that were
  identified in NOAA AR 10923. The solid green lines refer to the
  bisecting angles between the filaments P1 and P2. The red dotted
  contour has been drawn for LOS velocities greater than 2
  km~s$^{-1}$.}
\label{unsharp}
\end{figure}

\subsubsection{NOAA AR 11029}
\label{ar_29}
The continuum images and the red wing magnetograms of AR 11029 were
registered with the SP fast-mode map taken at 15:01-15:06 UT.  The
strong downflows observed at the umbra-penumbra boundary during this
time interval (top right panel of Figure~\ref{down_oct}) can be seen
in the following scan between 17:01-17:06 UT, nearly 2 hr later.  The
topmost downflowing zone is no longer visible, but the other velocity
patch remains at the same location. This strong downflowing blob
(indicated by the white arrow in the first scan) appears to rapidly
ingress towards the umbra-penumbra boundary with a speed of some
700~m~s$^{-1}$.

The scan taken between 19:01-19:06 UT demonstrates an intrusion of
bright photospheric granules into the spot, with the umbra slowly
being squeezed to occupy a much smaller area as depicted in the
continuum image. The downflows continue to be observed near the
shrinking umbral fragment. The scan made during 21:36-21:41 UT reveals
a drastic shrinkage of the umbra with only a small area of strong
downflows that appear close to where the umbra existed earlier (black
arrow in the continuum image of Figure~\ref{down_oct}). In comparison
to AR 10953, the area of the downflowing patch near the umbra-penumbra
boundary decreases with time. While the downflows occupy an area as
large as 6 arcsec$^2$ at 15:01 UT, this decreases to $\approx$2.3
arcsec$^2$ by 21:36 UT. Common characteristics of the downflows
observed in ARs 10953 and 11029 are their lifetimes of more than 6 hr
and the fact that the long-lived ones appear to be located near
the umbra-penumbra boundary. The more short-lived downflows are observed
elsewhere in the ARs.

\section{Discussion}
\label{orient}
Supersonic downflows associated with the Evershed flow are usually 
observed in the outer penumbra \citep{Josecarlos2001,Luis2004} 
or even beyond the sunspot boundary \citep{Valentin2009}. These 
downflows represent mass flux returning to the photosphere with
a polarity opposite to that of the sunspot. As they are seen in 
the outer penumbra they cannot explain the strong downflows we have detected
in the inner penumbra. One could suppose that the supersonic downflows
near the umbra-penumbra boundary are the photospheric manifestations of
some kind of inverse Evershed flow seen in the chromosphere. However, it is 
difficult to ascertain how such a chromospheric
phenomenon could produce supersonic downflows in the inner
penumbra at the photosphere.

Another mechanism that could perhaps explain the supersonic downflows 
is suggested below based on the orientation of the filaments in which they occur.
The continuum images were morphologically {\em opened}
\citep{Curto2008} and subtracted from their original image in order to
increase the contrast and isolate bright structures from the dark
background, filament crossings, and diffuse edges.  {\em Opening} is
the result of two operations namely {\em erosion} followed by {\em
  dilation}. These two operators look for the neighborhood minimum and
maximum where the search domain is defined by a structuring element
(SE). The size of the SE used for {\em opening} the images was
$7\times7$ pixels for the first two ARs while a $3\times3$ pixel SE
was sufficient for AR 11029. The above morphological operation was
chosen after discarding the Sobel and Roberts edge enhancement
operator, unsharp masking, and intensity thresholding. The results are
shown in Figure~\ref{unsharp} with reversed signs. It is observed that
in NOAA AR 10923, the strong downflows encompass two bisecting
filaments P1 and P2 and their likely point of intersection is marked
by a blue arrow. Such locations have been similarly marked for the
other ARs, although they are less obvious. The orientation of the
filaments resemble the post reconnection configuration illustrated in
Figure 5c of \citet{Ryutova2008a}, suggesting that the origin of the
downflows is the slingshot effect associated with the reconnection of
the filaments. The bisecting angles shown by the solid green lines in
AR 10923 were estimated to be $\approx$ 51$^{\circ}$ and 46$^{\circ}$
respectively. In AR 10953 only one half of the intersecting filaments
are visible, with the other end is possibly obscured by overlying
filaments.

\citet{Ryutova2008a} identified several instances of filaments
unwinding in a cork screw fashion leading to reconnection,
transient brightenings and twists in the penumbral 
filaments. This model was proposed as a possible mechanism to 
explain the existence of MJs. The magnetic configuration in 
the penumbra responsible for the MJs was recently investigated by
\citet{Magara2010}.  According to this work, MJs occur in the
intermediate region between nearly horizontal flux tubes and the
relatively vertical background field of the penumbra. A common aspect
of both models is that only parts and not the entire penumbral
filament participate in the reconnection, otherwise the filaments
would be destroyed after several reconnection events.

We have detected chromospheric enhancements at the position of the
strong downflows, although there is no strict one-to-one
correspondence. In the downflowing regions or in their immediate
vicinity there are also photospheric brightenings, with intensities
nearly the same as the quiet Sun.  This leads us to believe that a
slingshot reconnection may be responsible for the chromospheric
enhancements/jets, the photospheric brightenings, and the supersonic
downflows. However, the physical process is different from those
driving MJs, since their lifetimes are much shorter than those of the
events described here. While the lack of association of the strong
downflows at the umbra-penumbra boundary with the Evershed flow is
incontrovertible, the reconnection scenario has to be confirmed with
simultaneous high-cadence spectropolarimetric scans of the
photosphere.

\section{Summary}
\label{summary}
High resolution spectropolarimetric observations of 3 sunspots 
taken with {\em Hinode} indicate a new type of strong downflows 
in the inner penumbra. Such downflows are confined to isolated 
patches of 1.5--6 arcsec$^2$ in area. The LOS velocities were 
estimated using the SIR inversion code, retrieving supersonic or nearly 
supersonic values from 5 km~s$^{-1}$ to 10 km~s$^{-1}$ in the downflowing 
patches of all three ARs. The velocity patches occur very close to or 
at the umbra-penumbra boundary, which makes them the largest velocities 
ever detected at these locations in a sunspot. \citet{Shimizu2008}, on 
the other hand, reported frequent instances of supersonic downflows at 
the border of an umbra without a penumbra.

The strong velocities are usually associated with magnetic fields 
of more than 2 kG having the same polarity as the sunspot. This implies 
that the downflows are not driven by the Evershed flow, although the latter 
could independently be present at those locations. Intense and long lived 
chromospheric brightenings are seen near the strong photospheric downflows 
extending over an area of $\approx$ 1--2 arcsec$^2$. Photospheric 
brightenings nearly as intense as the quiet Sun are also present
in the downflowing regions or close to them. They are persistent 
as their chromospheric counterparts. In addition, the 
downflows may have lifetimes of up to 14 hr (or more) as can be judged from 
several consecutive Hinode/SP scans. 

The strong downflows are associated with penumbral filaments that appear 
to be twisted in the manner described by \citet{Ryutova2008b} that would 
arise from a reconnection process. Such a process is believed 
to produce the transient penumbral microjets. Microjets occur everywhere in 
the penumbra and the photospheric downflows observed with them are 
typically 1 km~s$^{-1}$, as reported by \citet{Jurcak2010}. The chromospheric
brightenings described in Section~\ref{ca-img} however, are more intense, long 
lived and bigger than the microjets but neither as intense nor intermittent 
as a flare.

The long-lived downflows reported here are not observed 
in all ARs as is evident from an inspection of the Level 1 SP 
maps\footnote{See {\em http://www.lmsal.com/solarsoft/data/hinode/sot/level1d}} 
from 2006 Nov to 2010 Feb. Although a total of 120 ARs were indexed 
by NOAA during this period, only 3 additional candidates were identified to
show strong downflows lasting 2 or 3 consecutive scans. These include 
AR 10956 on 2007 May 16, AR 10978 on 2007 Dec 12 and AR 10988 on 2008 April 1.

The strong photospheric downflows that we have described in this paper have not been 
reported earlier and as such represent a new phenomenon.
They are possibly driven by dynamic processes occurring in the inner penumbra 
which affect the chromosphere as well. The lack of a suitable theory for 
these events poses a challenge for future numerical simulations of sunspots.

\acknowledgments
Our sincere thanks to the {\em Hinode} team for providing the data
used in this paper and the referee for his/her comments. Hinode is 
a Japanese mission developed and launched by ISAS/JAXA, with NAOJ 
as domestic partner and NASA and STFC (UK) as international 
partners. It is operated by these agencies in co-operation with 
ESA and NSC (Norway). Financial support by the Spanish Ministerio 
de Ciencia e Innovaci\'on through project AYA2009-14105-C06-06 
as well as PCI2006-A7-0624 and by Junta de Andaluc\'{\i}a through project 
P07-TEP-2687 is gratefully acknowledged.

\end{document}